\documentclass{cimento}
\usepackage{graphics}
\usepackage{url}
\usepackage{color}
\usepackage{mathrsfs}
\usepackage{amssymb}
\usepackage{bm}
\usepackage[numbers]{natbib}
\usepackage{rotating}
\usepackage[normalem]{ulem}
\usepackage{epstopdf}
\usepackage{lipsum}
\usepackage{mathtools}
\usepackage{cuted}
\usepackage{enumitem}

\def\es0{$E_{\rm sym}(\rho_0)$}

\def\us0{$U_{\rm sym}(\rho_0,k_F)$~}

\def\l0{$L(\rho_0)$~}

\title{A theoretical overview of isospin and EOS effects in heavy-ion reactions at intermediate energies}
\author{Bao-An Li\from{ins:1}\thanks{Speaker's email address: Bao-An.Li@tamuc.edu}\ETC,
Bao-Jun Cai\from{ins:2},
Lie-Wen Chen\from{ins:3},
Wen-Jie Xie\from{ins:4},
Jun Xu\from{ins:5}
   \atque
Nai-Bo Zhang\from{ins:6}}
\instlist{
\inst{ins:1} Department of Physics and Astronomy, Texas A\&M
University-Commerce, Commerce, TX 75429-3011, USA
\inst{ins:2} Quantum Machine Learning Laboratory, Shadow Creator Inc., Shanghai 201208, China
\inst{ins:3} School of Physics and Astronomy, Shanghai Key Laboratory for
Particle Physics and Cosmology, and Key Laboratory for Particle Astrophysics and Cosmology (MOE),
Shanghai Jiao Tong University, Shanghai 200240, China
\inst{ins:4} Department of Physics, Yuncheng University, Yuncheng 044000, China
\inst{ins:5} Shanghai Advanced Research Institute, Chinese Academy of Sciences,
Shanghai 201210, and Shanghai Institute of Applied Physics, Chinese Academy
of Sciences, Shanghai 201800, China
\inst{ins:6} School of Physics, Southeast University, Nanjing 211189, China
  }

\begin{document}

\maketitle

\begin{abstract}
The isospin dependence of in-medium nuclear effective interactions is a fundamental issue in nuclear physics and has broad ramifications in astrophysics. Its uncertainties, especially the difference of neutron-proton interactions in the isosinglet and isotriplet channels, affect significantly the density and momentum dependence of the isovector single-nucleon potential and nucleon-nucleon short-range correlation in neutron-rich matter. Consequently, the neutron-proton effective mass splitting and the density dependence of nuclear symmetry energy are still rather uncertain. Heavy-ion reactions especially those involving rare isotopes is a useful tool for probing the isospin dependence of nuclear effective interactions through (1) the neutron-skin in coordinate and proton-skin in momentum of the initial state of colliding nuclei, (2) the density and momentum dependence of especially the isovector nuclear mean-field as well as (3) the isospin dependence of in-medium nucleon- nucleon cross sections. Observations of neutron stars especially since GW1710817 have also helped us significantly in understanding the isospin dependence of nuclear effective interactions. {\it We summarize here a review talk on these issues given at the 2021 International Workshop on multi-facets of EOS and Clustering. For details we refer the readers to the original publications and references therein}.
\end{abstract}

\section{The isospin-dependence of nuclear effective interactions and its ramifications on the EOS of neutron-rich matter}
The isospin degree of freedom and nuclear Equation of State (EOS) affect heavy-ion reactions at low-intermediate energies through both the in-medium nucleon-nucleon (NN) cross sections $\sigma^*_{NN}$ and the single-nucleon 
potential $U_{\tau}(k,\rho,\delta)$. The latter
for a nucleon ($\tau_3=\pm 1$ for neutrons/protons) of momentum $k$ in neutron-rich matter of density $\rho$ and isospin asymmetry $\delta=(\rho_n-\rho_p)/\rho$ can be written as \cite{Tesym}
\begin{equation}\label{sp}
U_{\tau}(k,\rho,\delta)=U_0(k,\rho)+\tau_3 U_{\rm sym,1}(k,\rho)\cdot\delta
+U_{\rm sym,2}(k,\rho)\cdot\delta^2+\tau_3 U_{\rm sym,3}(k,\rho)\cdot\delta^3+\mathcal{O}(\delta^4)
\end{equation}
in terms of the isoscalar $U_0(k,\rho)$ and $U_{\rm sym,2}(k,\rho)$ as well as the isovector $U_{\rm sym,1}(k,\rho)$ and $U_{\rm sym,3}(k,\rho)$ potentials, respectively.
While the energy per nucleon $E(\rho,\delta)$ in neutron-rich matter can be approximated as a parabolic function of $\delta$ as
$
E(\rho,\delta)=E_0(\rho)+E_{\rm{sym},2}(\rho )\delta ^{2} +\mathcal{O}(\delta^4)
$
in terms of the energy per nucleon $E_0(\rho)\equiv E(\rho,\delta=0)$ in symmetric nuclear matter (SNM) and the isospin-quadratic symmetry energy $E_{\rm{sym},2}(\rho )$ (denoted often by $E_{\rm{sym}}(\rho )$ or $S$ in the literature). The latter can be evaluated approximately from
$
E_{\rm{sym}}(\rho )\equiv E_{\rm{sym},2}(\rho )\equiv\left.\frac{1}{2}\frac{\partial ^{2}E(\rho,\delta )}{\partial \delta ^{2}}\right|_{\delta=0}\approx E(\rho,1)-E(\rho,0).
$
In terms of the components of the nucleon potentials, the symmetry energy can be written as \cite{Dab73c,XuC11}
\begin{equation}
E_{\rm{sym}}(\rho) 
=\frac{k_{\rm{F}}^2}{6 m^*_0(\rho,k_{\rm{F}})} +\frac{1}{2} U_{\rm{sym},1}(\rho,k_{\rm{F}}).
\label{FKW}
\end{equation}
where $k_F=(3\pi^2\rho/2)^{1/3}$ is the nucleon Fermi momentum and $m^*_0/m=(1+\frac{m}{\hbar^2k_{\rm F}}\partial U_0/\partial k)^{-1}|_{k_F}$ is the nucleon isoscalar effective mass of nucleons with a free mass $m$. In terms of the $m^*_0$, according to the Hugenholtz-Van Hove (HVH) theorem the slope $L(\rho) \equiv 3 \rho\frac{dE_{\rm sym}}{d\rho}$ of the symmetry energy at an arbitrary density $\rho$ is given by \cite{XuC11,Chen11}
\begin{equation}
L(\rho)= \frac{2}{3} \frac{\hbar^2 k_F^2}{2 m_0^*} + \frac{3}{2} U_{\rm sym,1}(\rho,k_F)
- \frac{1}{6}\Big(\frac{\hbar^2 k^3}{{m_0^*}^2}\frac{\partial m_0^*}{\partial k} \Big)|_{k_F}+\frac{\partial U_{\rm sym,1}}{\partial k}|_{k_F} k_F+ 3U_{\rm sym,2}(\rho,k_F). \label{Lexp2}
\end{equation}
The single-nucleon potential of Eq. (\ref{sp}) is a basic input in nuclear reaction models, e.g., various transport models for heavy-ion reactions. 
The above non-relativistic decompositions of both the symmetry energy $E_{\rm{sym}}(\rho)$ and its slope $L(\rho)$ in terms of the density and momentum dependences of the isoscalar and isovector single-nucleon potentials reveal clearly the underlying microscopic physics. The latter can be probed by using nuclear reactions at low-intermediate energies, especially those involving rare isotopes at radioactive beam facilities around the world. 

While the detailed density and momentum dependences of the single-nucleon potential in neutron-rich matter can be studied consistently in microscopic nuclear many-body theories, see, e.g., Ref. \cite{Rios-G}, to see the fundamental physics behind the $U_{\rm{sym},1}(\rho,k)$, it is  educational to simply look at its Hartree (direct) term at $k_F$ in the interacting Fermi gas model \cite{Xu-tensor}, namely, 
$
U_{\rm sym,1}(k_F,\rho)=
\frac{\rho}{4}\int [V_{T1}\cdot f^{T1}(r_{ij})-V_{T0}\cdot f^{T0}(r_{ij})]d^3r_{ij}
$
in terms of the isosinglet (T=0) and isotriplet (T=1) NN interactions $V_{T0}(r_{ij})$ and $V_{T1}(r_{ij})$, as well as the corresponding NN correlation functions $f^{T0}(r_{ij})$ and $f^{T1}(r_{ij})$, respectively. 
While $V_{nn}=V_{pp}=V_{np}$ in the T=1 channel due to the charge independence of NN interactions, 
the $V_{np}$ interactions and the associated NN correlations in the T=1 and T=0 channels are not the same due to the isospin dependence of the strong interactions. For example, it is well known that the tensor force and the resulting short-range correlation (SRC) in the T=0 channel is much stronger than that in the T=1 channel. However, several features of the SRC, such as its strength, the shape of the resulting high-momentum nucleon distributions in cold nuclear matter as well as their isospin and density dependence, are still poorly known. Obviously, these uncertainties will affect the density and momentum dependences of the single-nucleon potential, and thus the corresponding density dependence of the symmetry energy and the neutron-proton effective mass splitting. For a recent review on these issues, see, e.g., Ref. \cite{PPNP-Li}.  

\section{Probing the isospin-dependence of nuclear effective interactions by combining mutli-messengers from neutron star observations and nuclear reaction experiments}
Many studies in the literature over the last two decades have shown consistently that some observables of neutron stars and heavy-ion reactions bear clear signatures of the single-nucleon 
potential $U_{\tau}(k,\rho,\delta)$ as well as the corresponding symmetric nuclear matter (SNM) EOS $E_0(\rho)$ and the symmetry energy $E_{\rm{sym}}(\rho )$. To see intuitively, for example, how the latter can be extracted from combining information from both astrophysical observations and terrestrial nuclear experiments, we note that \cite{Li-Xie}
$
 E_{\textrm{sym}}(\rho)\approx E_{\textrm{sym}}(\rho_i)+\int_{\rho_i}^{\rho}\frac{P_{\rm{PNM}}(\rho_v)-P_{\rm{SNM}}(\rho_v)}{\rho_v^2}d\rho_v
$
where $\rho_i$ is a reference density. The pressure $P_{\rm{SNM}}$ in SNM over a broad density range has been extracted from terrestrial nuclear reaction experiments~\cite{Dan02}, while recent observations of neutron stars through multi-messengers have provided more information about the nuclear pressure in neutron-rich matter towards pure neutron matter (PNM) \cite{BALI19,BALI21}.  In the following, we present a few examples illustrating 
what we have learned recently from nuclear reactions and neutron stars about the isospin dependence of nuclear effective interactions in neutron-rich matter.
\\\\
\noindent{\bf Example1-Neutron-proton effective mass splitting in neutron-rich matter:}
\begin{figure}[htb]
\begin{center}
\vspace{0.65cm}
    \resizebox{1.\textwidth}{!}{
  \includegraphics{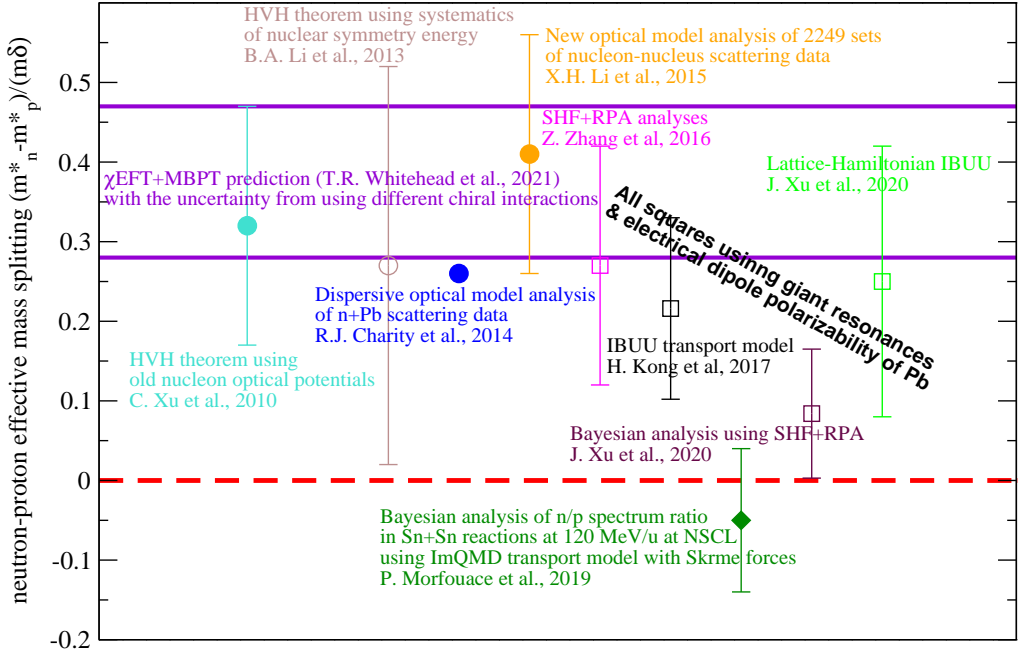}
  }
  \vspace{-0.1cm}
\caption{A survey of neutron-proton effective mass splitting at $\rho_0$ from analyzing nuclear reaction and structure experiments in comparison with the latest chiral effective field theory prediction (its 68\% confidence range is indicated by the horizontal violet lines).
}\label{emass}
\end{center}
\end{figure}
The momentum dependence of the single-nucleon potential is normally characterized by the nucleon effective masses $M^*_N$. 
The study on the effective masses of neutrons and protons in neutron-rich matter has a long history while many interesting issues remain to be resolved. In particular, due to our poor knowledge about the momentum dependence of isovector interaction, the isovector nucleon effective mass normally measured using the effective mass splitting $m^*_{\rm{n-p}}(\rho,\delta)\equiv(m_{\rm n}^*-m_{\rm p}^*)/m$ of neutrons and protons at the Fermi momentum $K_F$ has not been well constrained. The $m^*_{\rm{n-p}}(\rho,\delta)$ can be written as \cite{PPNP-Li}
$
m^*_{\rm{n-p}}(\rho,\delta)\approx 2\delta\frac{m}{k_{\rm{F}}}\left[-\frac{dU_{\rm{sym},1}}{dk}-\frac{k_{\rm{F}}}{3}\frac{d^2U_0}{dk^2}+\frac{1}{3}\frac{dU_0}{dk}\right]_{k_{\rm{F}}}\left(\frac{m^*_0}{m}\right)^2.
$
At saturation density, it is related to the  $E_{\rm{sym}}(\rho_0)$ and $L(\rho_0)$ via\,\cite{LiBA13}
\begin{equation}
m^*_{\rm{n-p}}(\rho_0,\delta)\approx\delta\cdot\frac{\displaystyle3E_{\rm{sym}}(\rho_0)-L(\rho_0)-3^{-1}({m}/{m^*_0})E_{\rm{F}}(\rho_0)}{\displaystyle
E_{\rm{F}}(\rho_0)\left({m}/{m_0^*}\right)^2}.\label{mnp}
\end{equation}
Therefore, the $m_{\rm n}^*$ is equal to, larger or smaller than the $m_{\rm p}^*$ depends on the symmetry energy and its slope. For example, with the empirical values of \es0=31 MeV, $m_0^*/m=0.7$ and $E_{\rm{F}}(\rho_0)=36$ MeV, a positive $m^*_{\rm{n-p}}(\rho_0,\delta)$ implies that the slope L should be less than $76$ MeV.  Interestingly, as we shall show in the next example, most of the extracted values of $L(\rho_0)$ from both terrestrial experiments and astrophysical observations satisfy this condition.

The nuclear physics community has devoted significant efforts to constraining the $m^*_{\rm{n-p}}(\rho,\delta)$ especially at the saturation density $\rho_0$. Shown in Fig. \ref{emass} is a survey of $m^*_{\rm{n-p}}/\delta$ 
at $\rho_0$. While there have been many, sometimes controversial, predictions on the $m^*_{\rm{n-p}}(\rho,\delta)$ using various nuclear many-body theories and interactions, only the latest prediction based on the many-body perturbation theory using chiral effective forces ($\chi$EFT+MBPT) by Whitehead et al \cite{Wh1,Wh2} is shown for a comparison. It is seen that most of the results from the indicated analyses are consistent with the 
$\chi$EFT+MBPT prediction within their 68\% error bars. In particular, all earlier analyses of nucleon-nucleus scattering data \cite{XuC10,Bob,XHLI15}, the analysis based on the HVH theorem using the 2013 systematics of nuclear symmetry energy \cite{LiBA13} as well as both the static and dynamical model analyses of the isoscalar and isovector giant resonances and the electrical dipole polarizability of $^{208}$Pb \cite{ZZ16,Kong17,Xu20,Xu20b} indicate surely a positive $m^*_{\rm{n-p}}(\rho_0,\delta)$. We notice that the four open squares are from using the same data set but different approaches. They give qualitatively consistent but quantitatively appreciably different $m^*_{\rm{n-p}}(\rho_0,\delta)$ values. One interesting exception is the result of $m^*_{\rm{n-p}}(\rho_0,\delta)=(-0.05\pm 0.09)\delta$ from a Bayesian analysis of the neutron/proton spectrum ratios in several Sn+Sn reactions at 120 MeV/nucleon at NSCL/MSU using an ImQMD transport model for nuclear reactions with Skyrme forces \cite{MSU}. A negative $m^*_{\rm{n-p}}(\rho_0,\delta)$ implies a characteristically different momentum dependence of the nucleon isovector potential inconsistent with the findings from optical model analyses of existing nucleon-nucleus scattering data \cite{PPNP-Li}.
\begin{figure}[htb]
\begin{center}
    \resizebox{1.\textwidth}{!}{
  \includegraphics{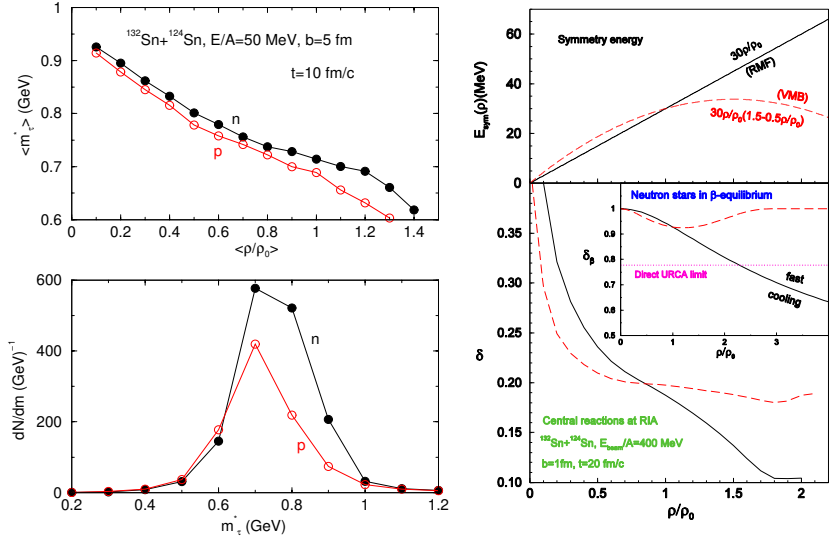}
  }
  \vspace{-0.1cm}
\caption{Left: The correlation between the average nucleon effective mass and the average nucleon density (top), and the
distribution of nucleon effective masses (bottom) in the reaction of $^{132}$Sn+$^{124}$ Sn at 10 fm/c with a beam energy of 50 MeV/A and an impact parameter of 5 fm \cite{Li-Chen05}. 
Right: The correlation between the isospin asymmetry and density during $^{132}$Sn+$^{124}$ Sn collisions at 20 fm/c with a beam energy of 400 MeV/A and an impact parameter of 1 fm and in the core of neutron stars at $\beta$-equilibrium (inset) \cite{IBUU2}.
}\label{ieffect}
\end{center}
\end{figure}

It is useful to note that heavy-ion reactions probe nucleon effective masses over a wide density range while nucleon-nucleus scatterings and/or giant resonances only probe those around $\rho_0$. The significant discrepancies between the results from analyzing the heavy-ion reaction data and those from other types of data present an interesting challenge. More studies especially with high energy radioactive beams may help address this issue. As an example, shown on the left in Fig. \ref{ieffect} are the correlation between the average nucleon effective mass and the average nucleon density (top), and the
distribution of nucleon effective masses (bottom) at 10 fm/c in the reaction of $^{132} ${\protect\small Sn+}$^{124}${\protect\small Sn} at a beam energy of 50 MeV/A
and an impact parameter of 5 fm \cite{Li-Chen05} within the IBUU transport model with a modified Gogny Hartree-Fock potential \cite{IBUU1}, while on the right are the correlations between the isospin asymmetries and densities in central $^{132} ${\protect\small Sn+}$^{124}${\protect\small Sn} collision at 400 MeV/A and neutron stars at $\beta$-equilibrium (inset) using the symmetry energy functionals shown in the upper-right frame \cite{IBUU2}. It is seen that during a typical heavy-ion reaction, the nucleon effective masses and the neutron-proton effective mass splitting depend on the density, and the matter formed can have various densities and isospin asymmetries depending on the symmetry energy functional used. 

A number of observables of heavy-ion reactions have been proposed as promising messengers of the underlying momentum dependence of the isovector potential and the corresponding $m^*_{\rm{n-p}}(\rho,\delta)$ \cite{LCK}. 
The reason is easily understandable. For example, the in-medium NN cross section $\sigma^*_{NN}$ is proportional to the square of the reduced effective mass of the two colliding nucleons. It will then affect the nuclear stopping power or the nucleon mean free path (MFP) $\lambda_{\rm{p}}^{-1}=\rho_{\rm{p}}\sigma_{\rm{pp}}^{\ast}+\rho_{\rm{n}}\sigma_{\rm{pn}}^{\ast}$ and $\lambda_{\rm{n}}^{-1}=\rho_{\rm{n}}\sigma_{\rm{nn}}^{\ast}+\rho_{\rm{p}}\sigma_{\rm{np}}^{\ast}$.
In terms of the nucleon effective k-mass $M_{J}^{\ast,\rm{k}}$, the nucleon MFP is given by \cite{Neg81}
$
\lambda_J=k_{\rm{R}}^J/(2M_{J}^{\ast,\rm{k}}|W_J|)
$
where $k_{\rm{R}}^J=[2M(E-U_J)]^{1/2}$, $U_J/W_J$ is the real/imaginary part
of the nucleon J's optical potential. 
Currently, no consensus on the $m^*_{\rm{n-p}}(\rho,\delta)$ has been reached from studying heavy-ion collisions as there are still large model dependences in both the transport models and the input single-nucleon potentials.
Nevertheless, it is encouraging to note that significant community efforts are being devoted to improving the situation \cite{Maria}. There are also strong experimental efforts to measure nuclear stopping powers and infer the underlying in-medium NN cross sections in neutron-rich matter, see, e.g., Ref. \cite{Ganil}.
\\\\
\noindent{\bf Example2-Progress in constraining nuclear symmetry energy around $(1-2)\rho_0$:}
\begin{figure}[htb]
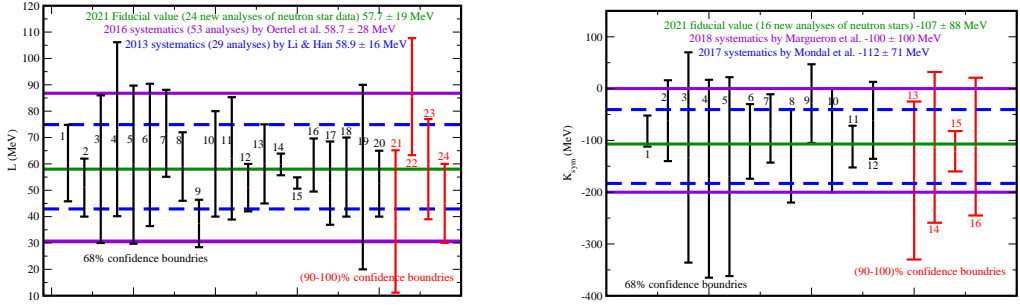

\begin{center}
\vspace{0.3cm}
  \resizebox{0.45\textwidth}{!}{
  \includegraphics{Lsym.eps}
  }
  \hspace{1cm}
    \resizebox{0.45\textwidth}{!}{
  \includegraphics{Ksym.eps}
  }
\caption{The slope parameter $L$ (left) and curvature parameter $K_{\rm{sym}}$ (right) of $E_{\rm{sym}}(\rho)$ at $\rho_0$ from analyses of neutron star observables since GW170817 in comparisons with the indicated earlier systematics. A complete list of individual analyses are given in Ref. \cite{BALI21} where these plots are taken from.}\label{LKsym}
\end{center}
\end{figure}
\begin{figure}[htb]
\begin{center}
  \resizebox{1.0\textwidth}{!}{
  \includegraphics{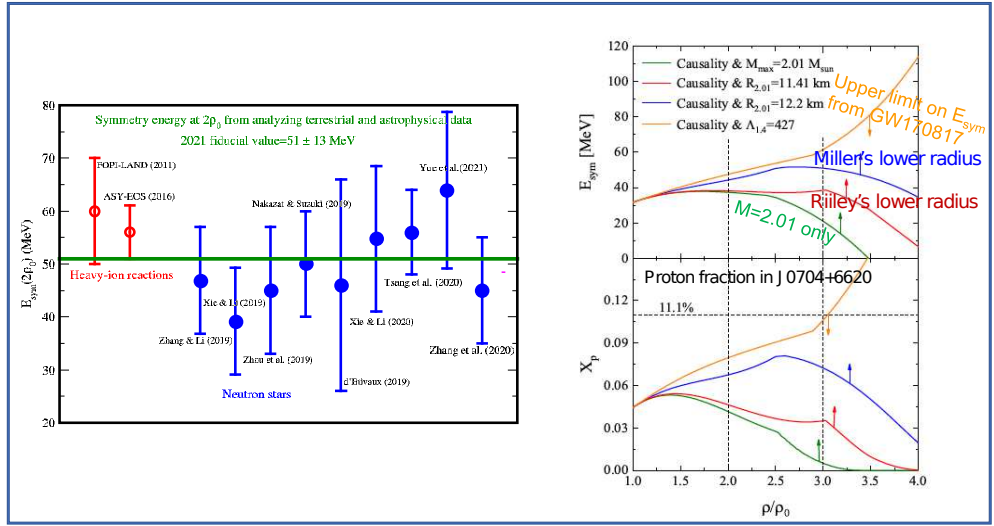}
  }
\caption{Left:  A survey of nuclear symmetry energy from heavy-ion reactions and neutron stars, taken from Ref. \cite{BALI21}. Right: constraints on the high-density symmetry energy and proton fraction using data from GW170817 and NICER+XMM-Newton's joint measurements of mass and radius for PSR J0740+6620, taken from Ref. \cite{NBZ-NICER}.}\label{Esym2}
\end{center}
\end{figure}
Much efforts have been made during the last two decades in constraining the $E_{\rm{sym}}(\rho)$. The latter can be characterized by its slope parameter $L$, curvature parameter $K_{\rm{sym}}$ and skewness parameter $J_{\rm{sym}}$ at $\rho_0$ according to 
$
E_{\rm{sym}}(\rho)\approx E_{\rm{sym}}(\rho_0)+L(\frac{\rho-\rho_0}{3\rho_0})+\frac{K_{\rm{sym}}}{2}(\frac{\rho-\rho_0}{3\rho_0})^2
+\frac{J_{\rm{sym}}}{6}(\frac{\rho-\rho_0}{3\rho_0})^3.
$
Significant progresses have been made in constraining the 
 $L$ and $K_{\rm{sym}}$ parameters using data from both terrestrial experiments and astrophysics observations, while the $J_{\rm{sym}}$ characterizing the stiffness of $E_{\rm{sym}}(\rho)$ around $(2-3)\rho_0$ remains unconstrained \cite{BALI21}. Shown in Fig. \ref{LKsym} are the slope parameter $L$ (left) and curvature parameter $K_{\rm{sym}}$ (right) from analyses of neutron star observables since GW170817 in comparisons with the indicated earlier systematics. A complete list of the individual analyses are given in Ref. \cite{BALI21}. It is seen that the new analyses of neutron star observables gave $L\approx 57.7\pm 19$ MeV and $K_{\rm{sym}}\approx -107\pm 88$ MeV in very good agreement with their systematics from earlier analyses. We notice that the extracted most probable L value from neutron star observables prefer a positive neutron-proton effective mass splitting according to Eq. (\ref{mnp}).

Recent neutron star observations have also led to some progresses in constraining the symmetry energy at suprasaturation densities. Shown on the left side of Fig.~\ref{Esym2} is a summary of symmetry energy at $2\rho_0$ from eleven analyses of heavy-ion reactions and neutron star properties in the literature. These analyses gave a mean value of $E_{\rm{sym}}(2\rho_0)\approx 51\pm 13$ MeV at 68\% confidence level as indicated by the green line. Obviously, more works are necessary to reduce the error bars. Most of the neutron star constraints are extracted from the radii and tidal deformations of canonical neutron stars. These observables are known to be sensitive mostly to pressures around $(1-2)\rho_0$ in neutron stars. Their constraints on the $E_{\rm{sym}}(\rho)$ around and above $2\rho_0$ are thus not so strong. Observables from more massive neutron stars were expected to place stronger constraints. To see how the NICER+XMM-Newton's simultaneous measurements of both the radius and mass of PSR J0740+6620 can help constrain the symmetry energy at densities above $2\rho_0$, shown in the upper-right is the extracted lower limits of  $E_{\rm{sym}}(\rho)$ when only the mass is observed (green line) or both the mass and radius are observed (red using the radius reported by Riley et al. \cite{Riley21} and blue using the radius reported by Miller et al \cite{Miller21}) from directly inverting the TOV equation within a 3-dimensional high-density EOS parameter space \cite{NBZ-NICER}. While the orange line is the upper limit of symmetry energy from analyzing the upper limit of tidal deformation of GW170817. The lower-right window shows the corresponding proton fractions in PSR J0740+6620. The constraining power of knowing both the mass and radius of this currently known most massive neutron star is clearly seen by comparing the green line with the red or blue line. While the difference between the red and blue lines indicates the systematic error from the two independent analyses of the same observational data.  Overall, while tighter constraints on the $E_{\rm{sym}}(\rho)$ around $(2-3)\rho_0$ have been obtained, there is still a strong need for better constraining the $E_{\rm{sym}}(\rho)$ at densities around and above $2\rho_0$ as indicated by the big gaps between the inferred upper and lower limits of $E_{\rm{sym}}(\rho)$. \\\\
\noindent{\bf Example3-The incompressibility, skewness and kurtosis of SNM EOS:}
The stiffness of SNM EOS can be characterized by its incompressibility $K_0$, skewness $J_0$ and kurtosis $Z_0$ parameters according to 
$
  E_{0}(\rho)\approx E_0(\rho_0)+\frac{K_0}{2}(\frac{\rho-\rho_0}{3\rho_0})^2+\frac{J_0}{6}(\frac{\rho-\rho_0}{3\rho_0})^3+\frac{Z_{0}}{24}(\frac{\rho-\rho_0}{3\rho_0})^4
$
with $E_0(\rho_0)$=-15.9 MeV. The corresponding pressure can be readily obtained from $P(\rho)=\rho^2\frac{dE_0(\rho)}{d\rho}$.
While the $K_0$ has been constrained to the range of 220 to 260 MeV according to various analyses of giant monopole resonances since about 1980 \cite{Bla1,Garg18}, the $J_0$ and $Z_0$ characterizing the stiffness of SNM EOS at super-saturation densities were only known roughly until very recently to be around $-200\pm 200$ and $-146\pm 1728$ MeV \cite{MM1}, respectively, mostly due to our poor knowledge about the three-body force in dense matter.
\begin{figure}[htb]
\begin{center}
  \resizebox{1.0\textwidth}{!}{
  \includegraphics{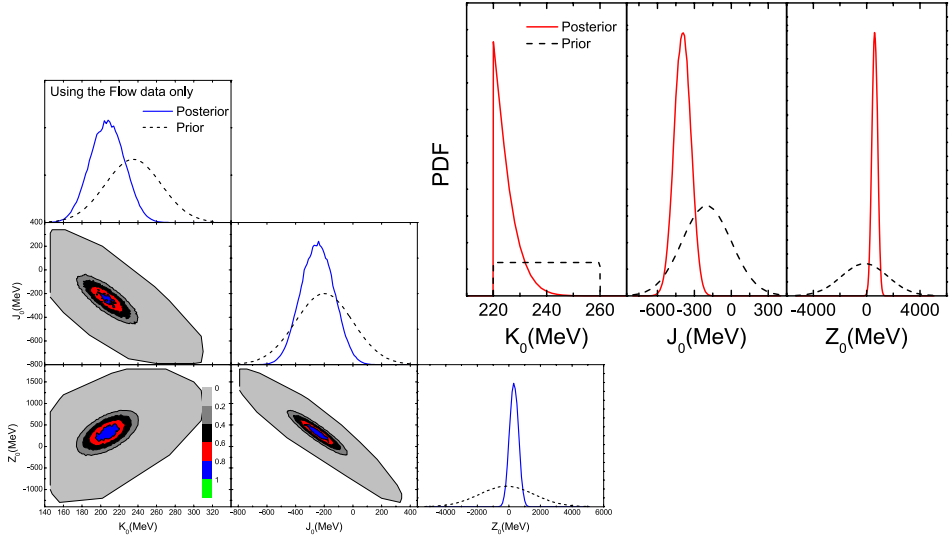}
  }
\caption{Left: The posterior PDFs for $K_0$, $J_0$ and $Z_0$ and their correlations obtained from the Bayesian analyses of the constraining bands on the SNM pressure using Gaussian priors. Right: using the uniform prior for 
$K_0$ in 220-260 MeV. Taken from Ref. \cite{Xie-JPG}.}\label{Pflow}
\end{center}
\end{figure}

Interestingly, some improvements were obtained very recently by a Bayesian calibration of the SNM EOS \cite{Xie-JPG} by using the constraining bands on the pressure in cold SNM in the density range of 1.3$\rho_0$ to 4.5$\rho_0$ from transport model analyses of kaon production and nuclear collective flow in relativistic heavy-ion collisions \cite{Dan02}.  
\begin{figure}[htb]
\begin{center}
  \resizebox{1.0\textwidth}{!}{
  \includegraphics{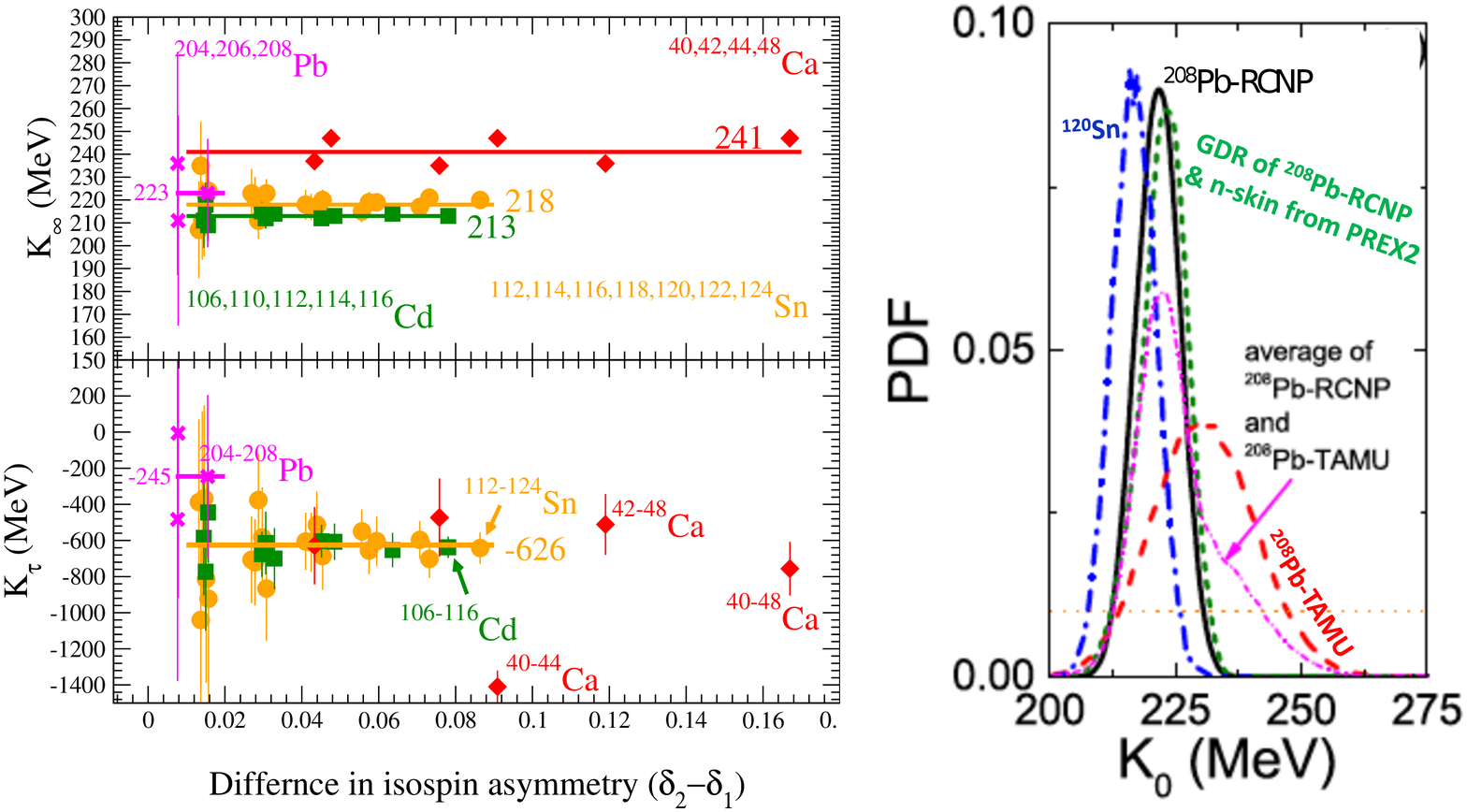}
  }
\caption{Left: $K_0$ and $K_{\tau}$ from a differential analysis of giant resonance data taken at RCNP, taken from Ref. \cite{Li21}. Right: the posterior PDF of $K_0$ from a Bayesian analysis of giant resonance data, taken from Ref. \cite{Jun21}. }\label{K0-new}
\end{center}
\end{figure}
As shown in Fig. \ref{Pflow}, assuming $K_0$, $J_0$ and $Z_0$ all have Gaussian priors centered around $235\pm 30$, $-200\pm 200$ and $-146\pm 1728$ MeV, respectively, the posterior most probable values of $J_0$ and $Z_0$ were found to be $J_0$=-240$^{+110}_{-120}$ MeV and $Z_0$=250$^{+300}_{-250}$ MeV, respectively, at 68\% confidence level. Their uncertainties are significantly smaller than their prior values. Interestingly, the inferred posterior PDF (probability distribution function) of $K_0$ peaks at 192 MeV, indicating that the flow data prefers a $K_0$ significantly smaller than its fiducial value from analyses of giant resonances. Moreover,  it was found that the posterior PDFs depend somewhat on the prior PDF used for the $K_0$. For example, shown in the right window of Fig.~\ref{Pflow} are the posterior PDFs of $K_0$, $J_0$ and $Z_0$ obtained by using a uniform prior for $K_0$ in 220-260 MeV while the prior PDFs for others are kept the same. Now, the posterior PDF of $K_0$ peaks at the lower boundary of its prior range, indicating again that the flow data prefers a smaller $K_0$ and the prior knowledge on $K_0$ from giant resonances is not sufficiently constraining. These results clearly indicate the needs for more stringent constraints on the $K_0$ before tighter constraints on the $J_0$ and $Z_0$ can be obtained. 

Partially due to the reasons mentioned above, some efforts have been made recently to infer the $K_0$ from the available world data of giant resonances in new ways significantly different from the traditional ``consensus approach" (i.e., the model must first reproduce all giant resonances and also other observables for finite nuclei before using it to predict the $K_0$ for SNM). One such effort is the simultaneous inference of both the $K_0$ and $K_{\tau}$ involved in the leptodermous expansion \cite{Bla1} of the incompressibility of finite nuclei $K_A \approx K_0(1+cA^{-1/3})+K_\tau \delta^2+K_{\rm {Cou}}Z^2A^{-4/3}$ in a differential approach \cite{Li21}. It uses the experimental $K_A$ values from RCNP for any two pairs in a given chain of isotopes. Shown in the left window of Fig. \ref{K0-new} are the $K_0$ and $K_{\tau}$ as functions of the difference ($\delta_2-\delta_1$) in isospin asymmetries of the isotope pairs used. The approach is not good for light nuclei for which the leptodermous expansion is less valid. While the inferred $K_0$ values from medium-heavy nuclei are well within the uncertainty range of its fiducial value, the inferred $K_{\tau}$ is much more precise than previously known \cite{Garg18}. It was also shown that the nuclear pairs with the largest separations in their isospin asymmetries give the most precise results. 
Another approach is the Bayesian inference using the SHF+RPA approach \cite{Jun21}. Compared to the traditional forward-modeling, this approach gives a more precisely quantified uncertainty estimation. Among the results shown in the right window of Fig. \ref{K0-new}, the posterior PDFs of $K_0$ from using the giant resonance data from RCNP for $^{208}$Pb is consistent with that from the differential approach shown on the left. Moreover, the slightly smaller $K_0$ values from analyzing the Sn data using both approaches are also consistent. However, considering the systematic errors from using the data from different laboratories and/or different nuclei, the $K_0$ values from these two new approaches are consistent but not more precise than its fiducial value. Thus, we have to work harder!

In summary, various isospin and EOS effects in nuclear reactions and properties of neutron stars enabled us to explore the isospin dependence of in-medium nuclear effective interactions. While significant progresses have been made by the community in constraining both the SNM EOS and nuclear symmetry energy below about $2\rho_0$, many interesting and challenging questions remain to be resolved.

\acknowledgments
This work was supported in part by the U.S. Department of Energy, Office of Science, under~Award Number DE-SC0013702, the CUSTIPEN (China-U.S. Theory Institute for Physics with Exotic Nuclei) under the U.S. Department of Energy Grant No. DE-SC0009971, the Yuncheng University Research Project under Grant No. YQ-2017005, the National Natural Science Foundation of China under Grant No. 11505150, No. 12005118, No. 11922514, No 11905302 and No. 11625521, National SKA Program of China Grant No. 2020SKA0120300, the Scientific and Technological Innovation Programs of Higher Education Institutions in Shanxi under Grant No. 2020L0550, and~the Shandong Provincial Natural Science Foundation under Grant No. ZR2020QA085.\\\\


\begin{thebibliography}{0}
\bibitem{Tesym} \BY{Li B. A., Ramos  \`{A}., Verde G., \atque Vida\~{n}a I. (eds.)}\\
  \IN{Topical Issue on Nuclear Symmetry Energy, Euro. Phys. J. A}{50}{2014}{9}.

\bibitem{Dab73c} \BY{Dabrowski J. \atque Haensel P.}
  \IN{Can. J. Phys.}{52}{1974}{1768}.

\bibitem{XuC11} \BY{Xu C., Li B. A., Chen L. W. \atque Ko C. M.}
  \IN{Nucl. Phys. A}{865}{2011}{1}.
  
 \bibitem{Chen11} \BY{Chen R.  \textit{et al.}}
 \IN{Phys. Rev. C}{85}{2012}{024305}.

\bibitem{Rios-G} \BY{Roshan Sellahewa \atque Arnau Rios}
  \IN{Phys. Rev. C}{90}{2014}{054327}.

\bibitem{Xu-tensor} \BY{Xu C. \atque Li B. A.}
  \IN{Phys. Rev. C}{81}{2010}{064612}.

\bibitem{PPNP-Li} \BY{Li B. A. \textit{et al.}}
  \IN{Prog. Part. Nucl. Phys.}{99}{2018}{29}.

\bibitem{Li-Xie} \BY{Li B. A. \atque Xie W. J.}
  \IN{Phys. Lett. B}{806}{2020}{135517}.

\bibitem{Dan02} \BY{Danielewicz P., Lacey R., \atque Lynch W. G.}
  \IN{Science}{298}{2002}{1592}.

\bibitem{BALI19} \BY{Li B. A. \textit{et al.}}
  \IN{Eur. Phys. J. A}{55}{2019}{117}.

\bibitem{BALI21} \BY{Li B. A. \textit{et al.}}
  \IN{Universe}{7}{2021}{182}.

\bibitem{LiBA13} \BY{Li B. A. \atque Han X.}
  \IN{Phys. Lett. B}{727}{2013}{276}.

\bibitem{Wh1} \BY{Whitehead T. R., Lim Y., \atque Holt J. W.}
  \IN{Phys. Rev. Lett.}{127}{2021}{182502}.

\bibitem{Wh2} \BY{Whitehead T. R.}
  private communications (2021).

\bibitem{XuC10} \BY{Xu C., Li B. A., \atque Chen L. W.}
  \IN{Phys. Rev. C}{82}{2010}{054607}.

\bibitem{Bob} \BY{Charity R. J. \textit{et al.}}
  \IN{Eur. Phys. J. A}{50}{2014}{23};
  Erratum: \SAME{50}{2014}{64}.

\bibitem{XHLI15} \BY{Li X. H. \textit{et al.}}
  \IN{Phys. Lett. B}{743}{2015}{408}.

\bibitem{ZZ16} \BY{Zhang Z. \atque Chen L. W.}
  \IN{Phys. Rev. C}{93}{2016}{034335}.

\bibitem{Kong17} \BY{Kong H. Y. \textit{et al.}}
  \IN{Phys. Rev. C}{95}{2017}{034324}.

\bibitem{Xu20} \BY{Xu J.\textit{et al.}}
  \IN{Phys. Lett. B}{810}{2020}{135820}.
  
\bibitem{Xu20b}\BY{Xu J. \atque Qin W. T.}
\IN{Phys. Rev. C} {102}{2020}{024306}.

\bibitem{MSU} \BY{Morfouace P. \textit{et al.}}
  \IN{Phys. Lett. B}{799}{2019}{135045}.
 
\bibitem{Li-Chen05} \BY{Li B. A. \atque Chen L. W.}
  \IN{Phys. Rev. C}{72}{2005}{064611}.
  
 \bibitem{IBUU1} \BY{Li B. A., Das C., Das Gupta S., \atque Gale C.}
  \IN{Nucl. Phys. A}{735}{2004}{563}.

\bibitem{IBUU2} \BY{Li B. A.}
  \IN{Phys. Rev. Lett.}{88}{2002}{192701}.


\bibitem{LCK} \BY{Li B. A., Chen L. W. \atque Ko C. M.}
\IN{Phys. Rep}{464}{2008}{113}.

\bibitem{Neg81} \BY{Negele J. W. \atque Yazaki K.}
  \IN{Phys. Rev. Lett.}{62}{1981}{71}.

\bibitem{Maria} \BY{Colonna M. \textit{et al.}}
  \IN{Phys. Rev. C}{104}{2021}{024603}.

\bibitem{Ganil} \BY{Lopez O. \textit{et al.}}
\IN{Phys. Rev. C}{90}{2014}{064602}. 

\bibitem{NBZ-NICER} \BY{Zhang N. B. \atque Li B. A.}
  \IN{Astrophys. J.}{921}{2021}{111}.

\bibitem{Riley21} \BY{Riley T. E. \textit{et al.}}
  \IN{Astrophys. J. Lett.}{918}{2021}{L27}.

\bibitem{Miller21} \BY{Miller M. C. \textit{et al.}}
  \IN{Astrophys. J. Lett.}{918}{2021}{L28}.

\bibitem{Bla1} \BY{Blaizot J. P.}
  \IN{Phys. Rep.}{64}{1980}{171}.

\bibitem{Garg18} \BY{Garg U. \atque Col\`{o} G.}
  \IN{Prog. Part. Nucl. Phys.}{101}{2018}{55}.

\bibitem{MM1} \BY{Margueron J., Hoffmann C. R., \atque Gulminelli F.}
  \IN{Phys. Rev. C}{97}{2018}{025805};
  \SAME{97}{2018}{025806}.

\bibitem{Xie-JPG} \BY{Xie W. J. \atque Li B. A.}
  \IN{J. Phys. G: Nucl. Part. Phys.}{48}{2021}{025110}.

\bibitem{Li21} \BY{Li B. A. \atque Xie W. J.}
  \IN{Phys. Rev. C}{104}{2021}{034610}.

\bibitem{Jun21} \BY{Xu J., Zhang Z., \atque Li B. A.}
  \IN{Phys. Rev. C}{104}{2021}{054324}.



\end{thebibliography}
\end{document}